# Observed Effect of Magnetic Fields on the Propagation of Magnetoacoustic Waves in the Lower Solar Atmosphere

J. K. Lawrence[1] and A. C. Cadavid[1]

[1]Department of Physics and Astronomy, California State University, Northridge,
18111 Nordhoff Street, Northridge, California 91330-8268, USA
(e-mail: *john.lawrence@csun.edu*, *ana.cadavid@csun.edu* )

**Abstract** We study *Hinode*/SOT-FG observations of intensity fluctuations in Ca II H-line and G-band image sequences and their relation to simultaneous and co-spatial magnetic field measurements. We explore the G-band and H-line intensity oscillation spectra both separately and comparatively via their relative phase differences, time delays and cross-coherences. In the non-magnetic situations, both sets of fluctuations show strong oscillatory power in the 3 – 7 mHz band centered at 4.5 mHz, but this is suppressed as magnetic field increases. A relative phase analysis gives a time delay of H-line after G-band of 20±1 s in non-magnetic situations implying a mean effective height difference of 140 km. The maximum coherence is at 4 – 7 mHz. Under strong magnetic influence the measured delay time shrinks to 11 s with the peak coherence near 4 mHz. A second coherence maximum appears between 7.5 – 10 mHz. Investigation of the locations of this doubled-frequency coherence locates it in diffuse rings outside photospheric magnetic structures. Some possible interpretations of these results are offered.

# 1. Introduction

To explore the hydromagnetic structure of the lower solar atmosphere, we investigated G-band and H-line intensity fluctuations on fine spatial scales (79 km pixel scale) and at high (0.35 – 0.59 min) cadences in filtergram sequences made by the Solar Optical Telescope (SOT-FG) on board the Hinode spacecraft. These two spectral bands are of interest for a variety of reasons. High-resolution and high-cadence G-band, H-line and magnetic image data are readily available from the *Hinode*/SOT-FG instrument. The G-band and H-line form over different height ranges in the lower solar atmosphere, and detailed response functions have been provided by Carlsson *et al*. (2007). The G-band and H-line data are commonly used as proxy data for magnetic fields (Muller and Roudier, 1984; Berger and Title, 2001; Rutten *et al*., 2001; de Wijn *et al*., 2009). Thus, of particular interest are the relationships of the G-band and H-line fluctuations, both separately and collectively, to the magnetic structure of the lower solar atmosphere. As a result, we use contemporaneous, co-spatial, line-of-sight magnetic field map sequences from SOT-FG as well as estimates of the heights in the atmosphere at which the Alfvén and sound speeds are equal. In most cases, though not all, the G-band and H-line intensity fluctuations at particular spatial locations have equal frequencies. This enables the study of such mutual properties as cross-coherence, phase shifts and time delays between the two sets of intensity fluctuations. In particular, we study these in terms of the importance of local magnetic fields to the dynamics and relate the results to models for wave propagation in the magnetized solar atmosphere.

In cases where magnetic effects are significant, a new second peak of coherence of the G-band and H-line intensity oscillations appears in the 7.5 – 10 mHz frequency band (see also Deubner and Fleck, 1990; Vecchio *et al.*, 2009). By mapping the spatial locations of cross-coherence in this frequency band, this effect can be associated with the edges of local magnetic field structures.

In Section 2 we describe the data and data preparation, including the calibration of the magnetic images and the potential-field extrapolation to heights above the observations. In Section 3 we study the power spectra of the H-line and G-band intensity fluctuations as a function of canopy height and local magnetic field strength respectively. In Section 4 the phase differences and time delays in the H-line and G-band signals are analyzed, and in Section 5 we investigate the coherence between the two signals. Additional discussion and conclusions are presented in Section 6.

## 2. Data

### 2.1 The *Hinode/*SOT-FG Data

The work presented in this paper is based on high-resolution, high-cadence time series of broadband filter imager (BFI) G-band and Ca II H-line images and narrowband filter imager (NFI) magnetograms of the Sun taken with *Hinode/*SOT-FG (Ichimoto *et al.*, 2008; Shimizu *et al.*, 2008; Suematsu *et al.*, 2008; Tsuneta *et al.*, 2008). The data were downloaded from the DARTS website and were processed to Level 1 with the SolarSoftware (SSW) FG-PREP package in IDL. The G-band (430.50 nm, band pass 0.8 nm) and Ca II H-line (396.85 nm, band pass 0.3 nm) SOT-FG images were 2 × 2 square-averaged to a pixel scale of 0.079 Mm. The magnetic data were shuttered FG *I* and *V* Fe I 630.2 nm, band pass 0.6 nm, *V/I* images with field of view 59 × 59 Mm and 2 × 2 binned to 0.115 Mm/pixel. The magnetograms were dilated and translated by us to co-register with the G-band and H-line filtergrams. We calibrated the SOT-FG magnetograms by comparison to a *Hinode/* SOT-SP (Spectro-Polarimeter) map made on 31 March 2007 that is both cospatial and contemporary with one of the data sets we analyzed. (The Community Spectro-polarimetric Analysis Center (CSAC) data are available at; http://www.csac.hao.ucar.edu). Histograms of magnetic field strengths yield strongly non-Gaussian distributions with heavy tails. Matching the positive and negative tails of histograms of the SP magnetic field values to those of suitably coarsened *Hinode*/SOT-FG data gave a conversion factor of $B/(V/I)$ = -24000±1000 G (gauss). This calibration was used in all data sets we studied.

In the following we exemplify our results using a data set taken on 14 April 2007 from UT 14:04:30 to UT 16:54:58 at a cadence of 21 s. In each camera cycle the H-line images were taken 3.2 s after the G-band, and then the magnetic data were taken 1.7 s after the H-line, or 4.9 s after the G-band. The useful field of view was 20 × 40 Mm, and the heliocentric cosine of this field was $\mu = 0.80$. Figures 1a and 1b display examples of G-band and H-line images. The results that we present in Figures 2, 4, and 5 have been checked against a sequence taken on 31 March 2007 from UT 18:26:31 to 23:46:58 at a cadence of 35.2 s. The H-line image was taken 3.2 s after the G-band and then the magnetogram 3.3 s after the H-line, or 6.5 s after the G-band. The useful field of view was 40 Mm x 40 Mm, and the heliocentric cosine was 0.93. Occasional salt-and pepper noise was handled by means of localized spatial median filtering. A few missing image gaps were interpolated in the time direction. Additional image sequences that contain both network and internetwork areas confirm the results given in Figure 6.

**2.2 Magnetic Fields and Magnetic Canopy**

To study the dependence of the G-band and H-line fluctuations on the local magnetic field we follow two approaches. We can either use the magnetic fields directly, or, in order to relate these effects more directly to the structure of the hydromagnetic solar atmosphere, we can translate the magnetic maps into maps of the height $Z_{EQ}$ at which the Alfvén and sound speeds are equal: $V_A = V_S$. This defines an "equipartition surface" or "canopy" below which the phenomena are dominated by plasma motions and above which the phenomena are dominated by magnetic effects (Cally, 2005). We note that the situation $V_A = V_S$ corresponds to plasma $\beta \equiv p_{gas}/p_{mag} = 1.2$ for a monatomic gas.

To find $Z_{EQ}$ at given locations we compare atmospheric gas pressure estimated from the Fontenla *et al.* (1990) model C atmosphere to upward potential field extrapolations of the magnetic field pressure (Alissandrakis, 1981). The atmospheric model sets the zero of $Z$ at the height where the opacity $\tau_{500} = 1$. However, the *Hinode* magnetic measurements are made using the magnetically sensitive Fe I 630.25 nm line. The core of this line forms at a height $\approx 0.25$ Mm above $\tau_{500} = 1$ (Alamanni *et al.*, 1990). However, the SOT-FG magnetic measurements involve shifts with polarization of the line wings which form lower in the atmosphere. We thus estimate the height of the magnetic measurements to be at $Z \approx 0.15$ Mm (Alamanni *et al.*, 1990; Jess *et al.*, 2010). Thus our calculations of the potential field extrapolation cover only $Z \geq 0.15$ Mm. Figures 1c and 1d respectively show an example of a line-of-sight magnetic image and the corresponding potential-field estimate of the equipartition $Z_{EQ}$.

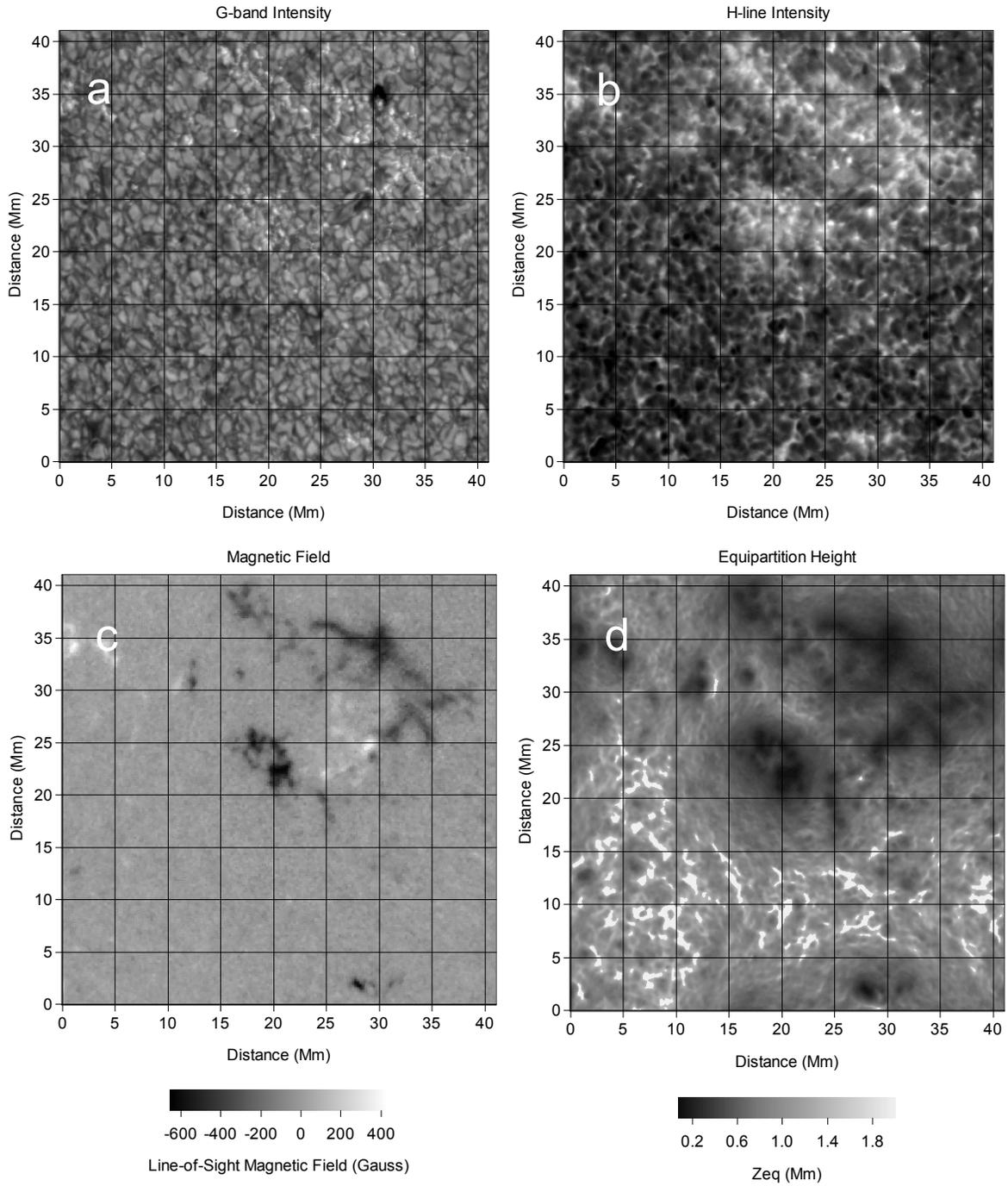

**Figure 1.** Snapshot images of the 14 April 2007 data set. (a) G-band intensity in arbitrary units, (b) H-line intensity in arbitrary units, (c) line-of-sight magnetic field in gauss. Panel (d) gives a potential-field estimate of the equipartition height $Z_{EQ}$ where the Alfvén and sound speeds are equal. In (d) $Z_{EQ} > 2$ Mm in the white areas. The pixel scale is 79 km.

**2.3 G-band and H-line Response Functions**

Carlsson *et al*. (2007) have calculated response functions for the *Hinode*/SOT-BFI Ca II H-line, and G-band filters that we are using here. The H-line response function peaks at ≈ 0.15 Mm, the mean height is at 0.247 Mm, and the response function extends upward beyond 1.0 Mm. The G-band response function, on the other hand, is concentrated significantly lower, with its peak near $Z = 0$ ($\tau_{500} = 1$) and a mean value of 0.074 Mm. About 2/3 of the H-line response function lies above our estimated effective height of the magnetic measurements at 0.15 Mm above $\tau_{500} = 1$, while only ≈ 1/5 of the G-band response function lies above. Because of this difference, in Section 3 we relate changes in the G-band oscillations to the value of the local magnetic field at $Z = 0.15$ Mm while relating changes in the H-line oscillations to the local values of the estimated canopy height $Z_{EQ} > 0.15$ Mm.

# 3. H-line and G-band Wavelet Spectra

**3.1 Morlet Wavelets**

The spatial locations of magnetic and intensity features drift with time. Therefore, a Fourier spectral analysis, which characterizes a single pixel column through the whole observational time span, will mix different physical regimes. Instead we use wavelet spectral analysis, which is localized in time and thus isolates the different regimes. The spectra are then characterized by either the local canopy height or the local magnetic field value whatever their spatial position.

The Morlet wavelet transforms were calculated with an IDL package due to Torrence and Compo (1998) http://paos.colorado.edu/research/wavelets. The Morlet wavelet mother function is a plane wave modulated by a Gaussian envelope: $\psi(\tau) = \pi^{-1/4} \exp(i\omega\tau) \exp(-\tau^2/2)$, where $\tau$ is the period. For the H-line case we take the wavelet power to be $P_H = |W(t, \tau)|^2/\tau$, where $W(t, \tau)$ is the wavelet transform at time $t$ with period $\tau$. In the G-band case there is an excess of long period correlation ("red noise") due to the slow evolution of photospheric features, so the power is then taken to be $P_G = |W(t, \tau)|^2/\tau^{3/2}$. See Lawrence *et al.* (2003) for a discussion of this point. Note that the definition $P_0 = |W(t, \tau)|^2$ gives a flat spectrum for white noise. In order to avoid spurious "end effects," we have calculated the wavelet transforms for the whole data sequence but have excluded power estimates near the beginning and the end of the data intervals.

In addition to the wavelet power, we will calculate the localized phases of oscillatory signals. The wavelet mother function is the sum of a real and an imaginary part: $\exp(i\omega\tau) = \cos(\omega\tau) + i \sin(\omega\tau)$. Then the local phase of $W(t, \tau)$ is $\Phi(t,\tau) =$ arctan [ Im($W(t,\tau)$) / Re($W(t,\tau)$) ]. We will use this to compare the phases of two signals in section 4.

**3.2 H-line and G-band Spectra**

Figure 2a shows the H-line spectra for the 14 April 2007 data set. These give spectral power per voxel, meaning that the power is divided by the number of space-time voxels contained in the corresponding $Z_{EQ}$ interval. We can see that for values of $Z_{EQ} \geq 0.8$ Mm there is a strong spectral feature in the frequency interval 3 mHz $< f <$ 7 mHz and centered at $f =$ 4.5 mHz . However, for $Z_{EQ} <$ 0.8 Mm the line weakens, and for $Z_{EQ} <$ 0.4 Mm it reaches a greatly reduced state with a small residual peak appearing at $f =$ 3.5 mHz.

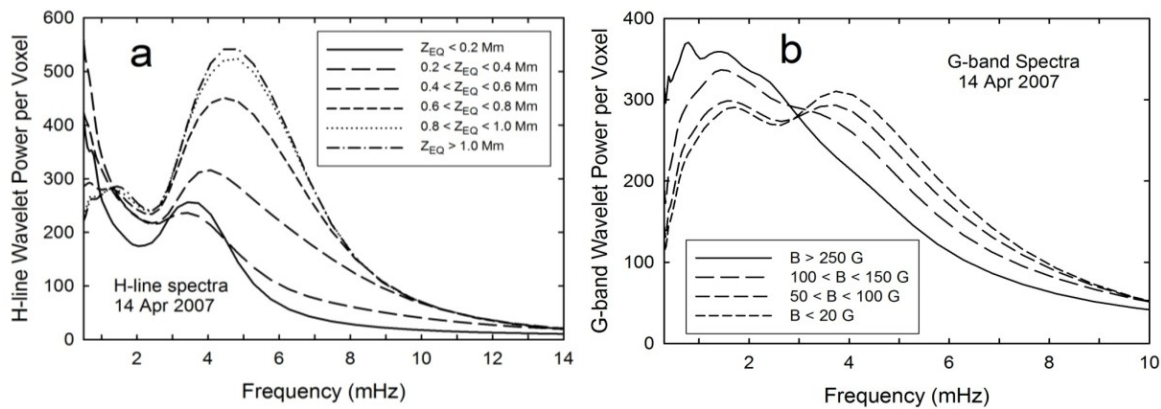

**Figure 2.** Morlet wavelet spectral power of (a) H-line intensity fluctuations per space-time voxel versus frequency in mHz for six intervals of canopy height $Z_{EQ}$ and (b) G-band intensity fluctuations per space-time voxel versus frequency in mHz for four intervals of local magnetic field strength. Both panels are from the 14 April 2007 data.

Figure 2b shows spectra of the 14 April 2007 G-band intensity fluctuations. For the reasons stated above these spectra are given according to varying intervals of local line-of-sight magnetic field ($B$) rather than canopy height. For the cases $B < 50$ G the spectra show an acoustic feature centered on a frequency $f = 4$ mHz. Also evident is a peak with characteristic time scale around 10 min (1.5 mHz) that may represent granular evolution. As $B$ rises above 50 G to 150 G the spectral peak at period 10 min is enhanced while the peak at 4 mHz disappears. The spectra for the H-line and G-band data of 31 March 2007 present similar results.

In order to interpret the results presented in Figure 2, we review some theoretical concepts regarding wave propagation in a magnetized environment (Cally, 2005, 2007; Shunker and Cally, 2006). In regions where $V_A < V_S$ the longitudinal acoustic wave corresponds to the fast mode while in the $V_A > V_S$ region it is a slow mode propagating along the magnetic field lines. As waves propagate upward across the equipartition height $V_A = V_S$, or $\beta = 1.2$, we say that mode "transmission" occurs when a fast acoustic (magnetic) wave changes to a slow acoustic (magnetic) wave. On the other hand mode "conversion" corresponds to the change from a fast (slow) acoustic wave to a fast (slow) magnetic wave. The process also depends upon the angle of inclination of the magnetic field with respect to the vertical, the "attack" angle of the wave vector with respect to the direction of the magnetic field, and the cut-off frequency for acoustic waves. Mode "transmission" is favored for small attack angles, low frequencies and a thin interaction region while mode "conversion" is favored for a large attack angle, high frequency and thicker interaction region.

The spectra in Figure 2 for both the H-line and G-band clearly show that oscillations with frequencies below the nominal acoustic cut-off (5.2 mHz) are being detected, and Figures 4 and 5 show that some propagation is taking place. McIntosh and Jefferies (2006) and Jefferies (2006) showed that these acoustic waves can leak through in regions in which the inclined magnetic field effectively lowers the acoustic cut-off, providing a guide for the propagating wave mode.

From Figure 2a it appears that when $Z_{EQ} > 0.8$ Mm then the H-line (and also the G-band) waves are seen below the magnetic canopy, so magnetic effects are not important, and the upwardly propagating waves are acoustic in nature. On the other hand, when $Z_{EQ} < 0.4$ Mm, the H-line spectra have settled into another state where now the H-line oscillations are seen above the canopy, though the G-band is mainly formed below. As the upwardly propagating acoustic waves cross the canopy interface they must undergo both mode transmission and conversion. Figure 2a shows that as the canopy height is progressively lowered from 0.8 Mm to 0.4 Mm an increasing proportion of upwardly propagating acoustic waves undergo mode conversion to a form that does not produce significant intensity fluctuations and is not detected, so the strong oscillatory peak is progressively removed. This, of course, corresponds to the magnetic "shadowing" of solar acoustic waves (Judge *et al*. 2001; Vecchio *et al*., 2007; Lawrence and Cadavid, 2010). The residual peak at frequency 3.5 mHz is compatible with the notion that wave transmission has been favored for this lower frequency.

In Figure 2b we see a similar effect. When the local magnetic field strength is relatively weak, *i.e.* less than 50 G, the G-band spectra are dominated by an oscillatory peak centered near 4 mHz and a secondary peak around 1.5 mHz possibly associated with granular evolution. As the field strength grows stronger, say, greater than 150 G, the oscillatory peak is suppressed, and the granular peak is enhanced. Because the G-band is formed so low in the atmosphere, it is hard to attribute this affect to mode conversion at an interface, but the effect is clearly magnetic since the power suppression comes into effect directly as the strength of the magnetic field is increased This also corresponds to a magnetic shadowing effect as observed in Lawrence and Cadavid, (2010).

## 4. Relative Phases and Time Delays

To investigate the relations between the G-band and H-line fluctuation time series, we calculated Morlet wavelet spectra of both data sets at each space-time voxel. Then, at each space-time voxel, we found the frequency of the maximum wavelet power in the G-band ($f_G$) and the H-line ($f_H$) spectra. When plotted as a two-dimensional histogram (Figure 3), the main concentrations of points lie along the diagonal line $f_G = f_H$ between 3 mHz and 6 mHz. Thus the G-band and H-line intensity fluctuations mostly have equal frequencies in the ranges corresponding to the strong spectral features seen in Figures 2a and b.

For equal frequencies it is of interest to investigate the relative phases of the fluctuations. At each frequency we produce a histogram counting the number of occurrences of a given phase difference in 5° bins from -180° to +180°. The results for the 14 April data for the high $Z_{EQ}$ (non-magnetic) situation are shown in Figure 4a, and for low $Z_{EQ}$ (magnetic) situation in Figure 5a.

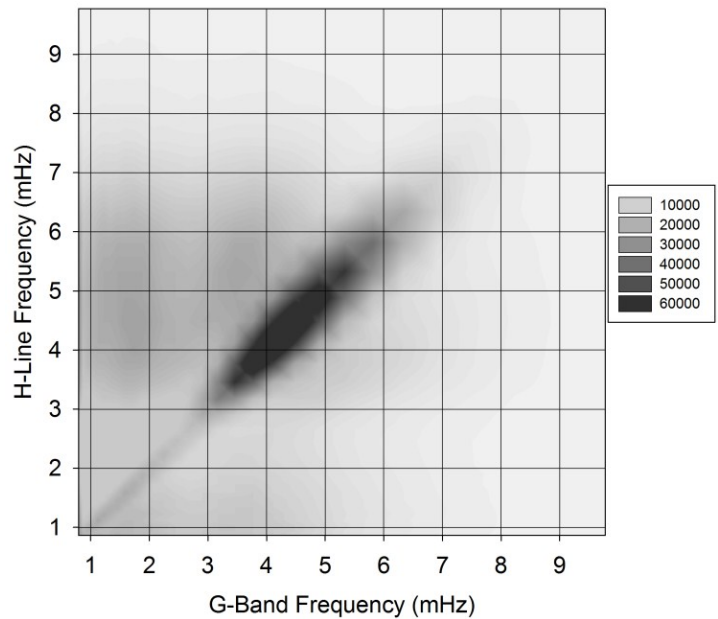

**Figure 3.** Histogram of pairs of H-line and G-band maximum spectral frequencies calculated at all space-time voxels in the data sets. The strongest concentration appears for equal spectral maxima at frequencies between 3 and 6 mHz.

In Figure 4a ($Z_{EQ} > 0.8$ Mm) it is clear that at particular frequencies, especially around $f \approx 5$ mHz, the occurrences of phase shifts are concentrated strongly around a particular maximum value. This indicates relatively high correlation between the G-band and H-line signals near that frequency.

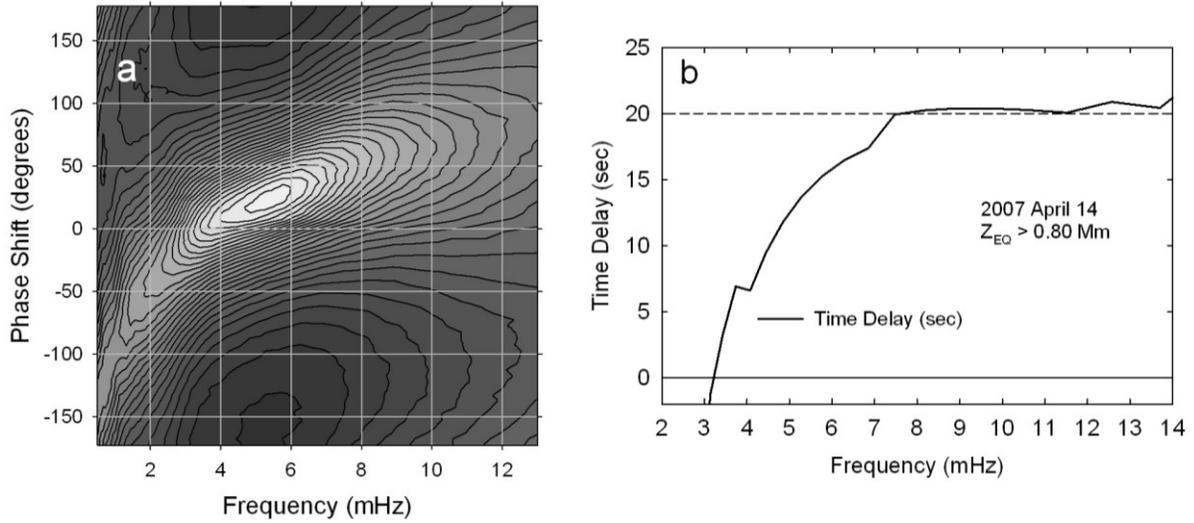

**Figure 4** (a) Histogram of the number of occurrences of particular phase shifts between the H-line and G-band wavelets plotted versus phase shift in degrees and frequency in mHz. (b) The curve indicates time delays in seconds derived from the maxima along the ridgeline in (a). These are constructed from the 14 April 2007 data at locations where the canopy height $Z_{EQ} > 0.80$ Mm.

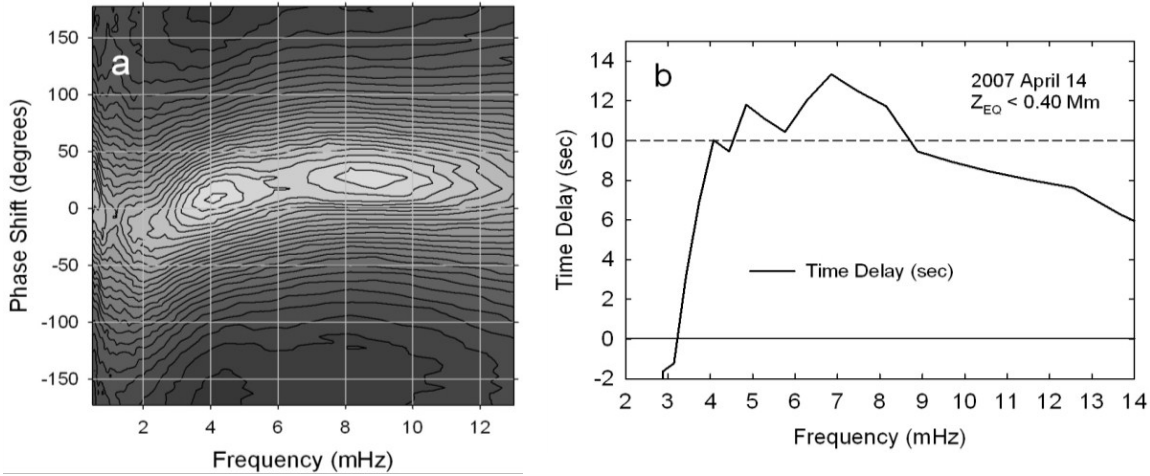

**Figure 5.** (a) Histogram of the number of occurrences of particular phase shifts between the H-line and G-band wavelets plotted versus phase shift in degrees and frequency in mHz. (b) The curve indicates time delays in seconds derived from the maxima along the ridgeline in (a). These are constructed from the 14 April 2007 data at locations where the canopy height $Z_{EQ} < 0.40$ Mm.

Figure 5 presents the results for $Z_{EQ} < 0.4$ Mm which corresponds to the more magnetic regions of the data set. In Figure 5a the low frequency limit of the phase differences in the April 14 2007 data now are closer to $\Delta\Phi \approx 0$, reflecting the fact that in the presence of magnetic flux elements both the G-band and the H-line intensities tend to be bright. As seen clearly in Figure 5a, there are now two maxima of the histogram. One maximum occurs around $f \approx 4.5$ mHz and may represent a shift of the cutoff frequency from 5 to 4.5 mHz. The new, broader peak lies around $f \approx 7.5 - 10$ mHz. For the March 31 2007 data the two maxima are even more distinct. In the low $Z_{EQ}$ cases the correlation is generally weaker than in the high $Z_{EQ}$ cases, probably because the magnetic solar atmosphere is less homogeneous than the quiet Sun atmosphere, and the phase shifts are more varied.

The important point here is that there are two peaks in Figure 5a. Figures 4a and 5a are histograms that show how the results of a large number of measurements of the phase difference are distributed in wave frequency and angle. We can estimate the 1σ error in a histogram value of $N$ as $\sqrt{N}$. The two peaks in Figure 5a have $N \approx 44700$, and the valley between has $N \approx 43700$. Thus, the difference of $\approx 1000$ between peaks and valley has an rms 1σ uncertainty of $\approx 300$. So the existence of the second peak is a 3σ result and is significant to $\approx 99\%$ confidence. In addition, we have examined other data sets and found a similar result. Previous authors have seen it in other data sources.
Vecchio *et al*. (2009) using the photospheric Fe I line at 709.0 nm and the chromospheric Ca II 854.2nm line obtained with IBIS, found that the coherence spectra for velocity signals showed a maximum at around 7mHz. In earlier work Deubner and Fleck (1990) had identified a similar feature in the coherence spectra for Fe 849.6 and Ca II velocity signals. Vecchio *et al*. (2009) suggest that this could be connected to the "aureoles" of 6 mHz enhanced power found near network elements and active regions (*e.g.* Krijger *et al*., 2001). In other work Andić (2008) found an increase in the coherence between intensity signals in the lines Fe I 543.29 nm and Fe I 543.35 nm and for frequencies 9 - 12mHz. In Figures 4b and 5b the curves are constructed by finding the maxima of the phase difference histograms in Figures 4a and 5a at each frequency value. These were estimated from cross sectional plots with allowance for the effects of noise. The time delay, in seconds, between the G-band and H-line signals is given by $\Delta t = \Delta\Phi/0.36f$, where $\Delta\Phi$ is the phase difference in degrees and $f$ is the frequency in mHz. In Figures 4b and 5b a positive value means that the H-line trails the G-band by $\Delta t$. Note that for the 14 April 2007 data we add an additional 3.2 s because of the time delay between the two images in the spacecraft camera system. The corresponding correction is applied to the other data set.

In the limit of low frequency the phase differences in Figure 4a approach -180°. This reflects the fact that on time scales of photospheric evolution the G-band and H-line intensities have opposite spatial distributions: The G-band images are darker in the intergranular lanes while the H-line images are brighter, and the reverse in the granular interiors (Evans and Catalano, 1972; Suemoto, Hiei, and Nakagomi, 1987, 1990). In Figure 4b, as the frequency increases to just above 3 mHz the time delay crosses zero, indicating a standing wave. Then as the frequency increases from 3.5 mHz to about 7 mHz the time delay increases indicating increasing upward propagation. Notably, this frequency range coincides with the highest peak in the histogram of Figure 4a. Then, for frequencies above 6 or 7 mHz the time delay takes on a constant value $20 \pm 1$ s. At a sound speed of 7 km s$^{-1}$ this time delay implies a height difference $\Delta Z = 0.14$ Mm in agreement with the *Hinode* response functions presented by Carlsson *et al.* (2007). The 31 March 2007 data present a similar result with a time delay of $22 \pm 1$ s.

As was so for the high $Z_{EQ}$ cases, the time delays in the low $Z_{EQ}$ cases pass through zero just above $f = 3$ mHz. They then increase quickly to values near 12 s for 4 mHz $< f <$ 8 mHz and then slowly decrease for higher frequencies. The reduced time delays in the magnetic cases may imply an increased wave speed or a change in the spatial geometry. Given the cadences of relevant Hinode observations ($\geq 21$ s) we cannot directly calculate a time delay around 12s and so must rely on phase techniques. It is worth noting that by using ground-based data at a cadence of ≈4.2s Lawrence *et al.* (2011) find a negligible time delay between a G-band and a K-line signal.

## 5. G-band and H-line Coherences

Based on the results from the previous section we further investigated the spatial distribution of sites of high-frequency G-band – H-line coherency. We calculated wavelet cross-coherence spectra (Torrence and Webster, 1999) for the 14 April data and used these to localize the enhanced coherence in the 7.5 – 10 mHz frequency band that is indicated by the second peak in the histogram of Figure 5b. We then time-averaged all the cross-coherences in this frequency band to produce the image shown in Figure 6a. Comparison to a time-averaged line-of-sight magnetic field map (Figure 6b) shows that the enhanced coherence in the frequency band 7.5-10 mHz is localized around the borders of magnetic elements. At the same time, at the actual locations of magnetic fields the coherence is suppressed. This supports the connection with the "halos" or "aureoles" of enhanced power noted by Krijger *et al*. (2001), even if the characteristic frequencies are different. Additional image sequences that contain both network and internetwork areas confirm the results given in Figure 6.

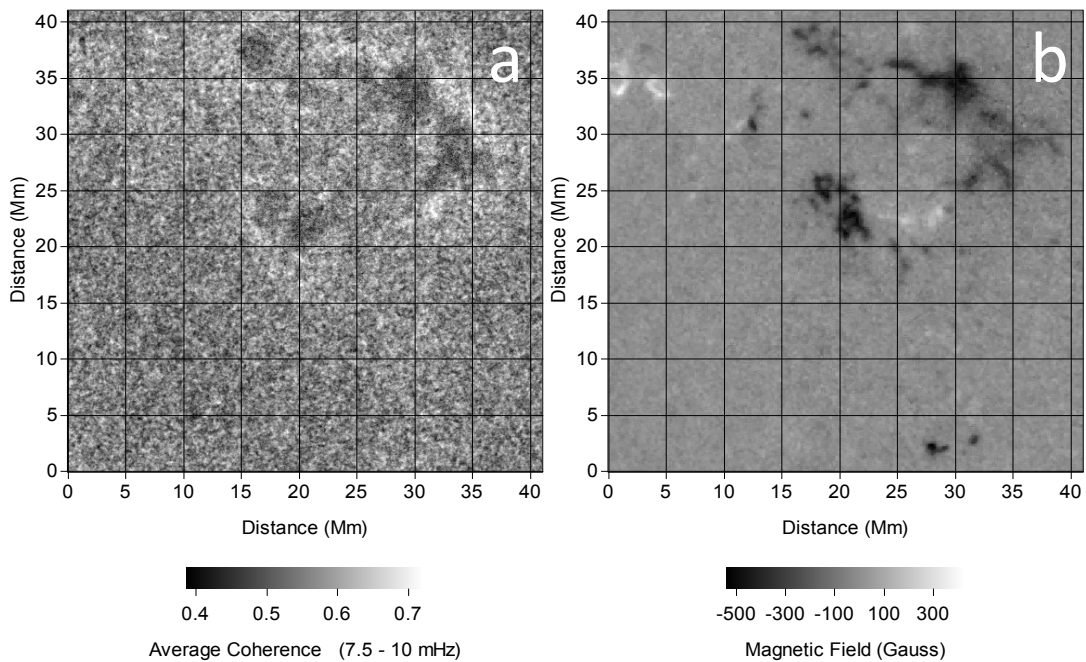

**Figure 6.** (a) Map of the time-averaged spatial location of enhanced G-band to H-line coherency in the 7.5 to 10 mHz frequency band for the 14 Apr 2007 data set. (b) Time-averaged line-of-sight magnetic field in the 14 April 2007 data set.

## 6. Discussion and Conclusions

The time differences at about 20 s between the G-band and H-line propagating wave signals in the non-magnetic case, imply a height difference $\Delta Z = 0.14$ Mm for an acoustic speed of 7 km s$^{-1}$. This is in agreement with the Hinode response functions presented by Carlsson *et al*. (2007) and is compatible with the results encountered through the power spectrum analysis. The latter established that for the non-magnetic regions both the G-band and H-line are formed below the $Z_{EQ}$ and the acoustic longitudinal wave propagates basically undisturbed. Lawrence *et al.* (2011) using superposed epoch analysis found a time delay of 21 s between G-band and K-line signals in non-magnetic regions, which further confirms the present result. The lesser time delay in the magnetic case may possibly imply either an increased wave speed or a change in the spatial geometry. For this case Lawrence *et al.* (2011) found a negligibly short time delay between the G-band and K-line signals.

In the non-magnetic situation shown in Figure 4a we find strongly enhanced coherence of the G-band and H-line oscillations that extends roughly from 3 – 8 mHz and is peaked near 5.2 mHz, the nominal cutoff frequency for wave propagation. In the strongly magnetic case of Figure 5a we find lower coherence that is peaked near 4.5 mHz and now a secondary, frequency-doubled, maximum peaked around 9 mHz.

The appearance of a secondary, high-frequency, maximum in the coherence in the magnetically influenced case of Figure 5 deserves further exploration. A map of the cross-coherence in the 4 – 5 mHz band, which includes a strong admixture of standing waves, is spread uniformly over the whole image (not shown). As described above, we produced a map of the spatial location of enhanced G/H cross-coherence in the 7.5 – 10 mHz band for the 14 April 2007 data set. Comparison of this map, given in Figure 6a, to the corresponding magnetic image in Figure 6b gives some confirmation to the suspected connection (Vecchio *et al*., 2009) to aureoles of enhanced oscillatory power around strong magnetic features (Krijger *et al*. (2001) and several references therein). The features seen in Figure 6a, however, are different in that they indicate enhanced high-frequency G-band – H-line cross-coherence rather than wave power. The two sets of oscillations appear to communicate with each other more readily at these frequencies and at these locations, perhaps by way of a resonance effect across the magnetic canopy.

We make a distinction between our association of the enhanced cross-coherence with inclined magnetic fields, on the one hand, and the phenomenon of "leaking" of solar acoustic waves below the acoustic cutoff frequency (< 5 mHz) that also is associated with inclined fields (*e.g.* Jefferies *et al*., 2006), on the other. In our case the enhanced cross-correlation takes place in a relatively distinct doubled-frequency band of 7.5 – 10 mHz.

A scenario that can encompass both the decreased phase travel time and the enhanced cross-correlation around the locations of magnetic features is offered in a recent publication by Nutto *et al*. (2012). This is based on numerical simulations of the lower solar atmosphere in which acoustic waves are emitted near the feet of magnetic elements. These waves propagate upward into the region where the Alfvén velocity $V_A$ is greater than the sound speed $V_S$, a level corresponding to $β < 1.2$. When passing upward through this level, the waves undergo a mode conversion from the fast acoustic mode to the fast magnetic mode. Above this, gradients in the value of $V_A$ cause the fast magnetic waves to refract outwards from the magnetic element and then downward again. Around the highest part of the wave path, the waves are propagating nearly horizontally, and the wave fronts are oriented vertically over large ranges of height. As a result, to the extent that both the G-band and H-line intensity fluctuations can be seen above the canopy, the measurement of the wave amplitudes at different heights in this region would indicate a small or zero vertical phase travel time. Likewise, there would be a stronger correlation between them.

The phenomenon of frequency doubling of plasma waves has appeared before in theoretical literature (see Jess *et al.*, 2012). Ulmschneider, Zähringer, and Musielak (1991) simulated the properties of nonlinear transverse waves propagating upward through the solar atmosphere along a thin magnetic flux tube. It was found that a longitudinal wave was generated at double the transverse frequency. This would be compatible with the scenario presented by Nutto *et al.* (2012) for both the signals measured above the canopy therefore in a low β region. In an alternative scenario, a theoretical calculation of the propagation of MHD waves in a high β plasma, Song, Russell, and Chen (1998) found that for transverse magnetic waves propagating along the magnetic field a compressible mode develops with twice the frequency of the primary wave. Such waves might develop in the high β region below the canopy in the lower solar atmosphere, and then pass via mode transmission across the canopy. Both of the above mechanisms rely on non-linearities in the plasma properties and imply the presence of turbulence, shown to be present at the locations of magnetic features (Lawrence *et al.*, 2011).

**Acknowledgments** *Hinode* is a Japanese mission developed and launched by ISAS/JAXA, collaborating with NAOJ as a domestic partner, and NASA and STFC (UK) as international partners. Scientific operation of the Hinode mission is conducted by the Hinode science team organized at ISAS/JAXA. This team mainly consists of scientists from institutes in the partner countries. Support for the post-launch operation is provided by JAXA and NAOJ (Japan), STFC (U.K.), NASA, ESA, and NSC (Norway). The authors are grateful to Dr. Thomas E. Berger for helpful discussions of the *Hinode*/SOT – FG data. Hinode SOT/SP Inversions were conducted at NCAR under the framework of the Community Spectro-polarimetric Analysis Center http://www.csac.hao.ucar.edu. Wavelet software was provided by C. Torrence and G. Compo and is available at http://paos.colorado.edu/research/wavelets/.

# References

Alamanni, N., Bertello, L., Cavallini, F., Ceppatelli, G., Righini, A.: 1990, *Astron. Astrophys.* **231**, 518.

Alissandrakis, C.E.: 1981, *Astron. Astrophys.* **100**, 197.

Andić, A.: 2008, *Serb. Astron. J.* **177**, 87.

Berger, T.E., Title, A.M.: 2001, *Astrophys. J.* **553**, 449.

Cally, P.S.:2005, *Mon. Not. Roy. Astron. Soc*. **358**, 353.

Cally, P.S.:2007, *Astron. Nachr.* **328**, 286.

Carlsson, M., Hansteen, V.H., De Pontieu, B., McIntosh, S., Tarbell, T.D., Shine, R., *et al.*: 2007, *Publ. Astron. Soc. Japan* **59**, S663.

Deubner, F.-L., Fleck, B.: 1990, *Astron. Astrophys.* **228**, 506.

De Wijn, A.G., Stenflo, J.O., Solanki, S.K., Tsuneta, S.: 2009, *Space Sci. Rev.* **144**, 274.

Evans, J. W., Catalano, C. P.: 1972, *Solar Phys*. **27**, 299.

Ichimoto, K., Lites, B., Elmore, D., Suematsu, Y., Tsuneta, S., Katsukawa, Y., *et al.*: 2008, *Solar Phys.* **249**, 233.

Jefferies, S.M., McIntosh, S.W., Armstrong, J.D., Bogdan, T.J., Cacciani, A., Fleck, B.: 2006, *Astrophys. J. Lett.* **648**, L151.

Jess, D. B., Mathioudakis, M., Christian, D.J., Keenan, F.P., Ryans, R.S.I., Crockett, P.J.: 2010, *Solar Phys*. **261**, 363.

Jess, D.B., Pascoe, D.J., Christian, D.J., Mathioudakis, M., Keys, P.H., Keenan, F.P.: 2012, *Astrophys. J. Lett*. **744**, L5.

Judge, P. G., Tarbell, T. D., Wilhelm, K.: 2001, *Astrophys. J.* **554**, 424.

Krijger, J.M., Rutten, R.J., Lites, B.W., Straus, Th., Shine, R.A., Tarbell, T.D.: 2001, *Astron. Astrophys.* **379**, 1052.

Lawrence, J.K., Cadavid, A.C., Miccolis, D., Berger, T.E., Ruzmaikin, A.: 2003, *Astrophys. J.* **597**, 1178.

Lawrence, J.K., Cadavid, A.C.: 2010, *Solar Phys.* **261**, 35.

Lawrence, J.K., Cadavid, A.C., Christian D.J., Jess, D.B., Mathioudakis, M.: 2011, *Astrophys. J. Lett*. **743**, L24.

McIntosh, S.W., Jefferies, S.M.: 2006, *Astrophys. J. Lett.* **647**, L77.

Muller, R., Roudier, T.: 1984, *Solar Phys.* **94**, 33.

Nagashima, K., Sekii, T., Kosovichev, A.G., Shibahashi, H., Tsuneta, S., Ichimoto, K., *et al.*: 2007, *Publ. Astron. Soc. Japan* **59**, S636.

Nutto, C., Steiner, O., Schaffenberger, W., Roth, M.: 2012, *Astron. Astrophys.* **538**, A78.

Rutten, R.J., Hammerschlag, R.H., Sütterlin, P., Bettonvil, F.C.M.: 2001, In: Sigwarth, M. (ed), *Advanced Solar Polarimetry -- Theory, Observation, and Instrumentation*, *ASP Conf. Ser.* **236**, 25.

Schunker, H., Cally, P.S.:2006, *Mon. Not. Roy. Astron. Soc.* **372**, 551.

Shimizu, T., Nagata, S., Tsuneta, S., Tarbell, T., Edwards, C., Shine, R., *et al.*: 2008, *Solar Phys.* **249**, 221.

Song, P., Russell, C.T., Chen, L.: 1998, *J. Geophys. Res.* **103**, 29569.

Suematsu, Y., Tsuneta, S., Ichimoto, K., Shimizu, T., Otsubo, M., Katsukawa, Y., *et al.*: 2008, *Solar Phys.* **249**, 197.

Suemoto, Z., Hiei, E., Nakagomi, Y.: 1987, *Solar. Phys.* **112**, 59.

Suemoto, Z., Hiei, E., Nakagomi, Y.: 1990, *Solar. Phys.* **127**, 11.

Torrence, C., Compo, G.P.: 1998, *Bull. Amer. Meteor. Soc.* **79**, 61.

Torrence, C., Webster, P.J.: 1999, *J. Climate* **12**, 2679.

Tsuneta, S., Suematsu, Y., Ichimoto, K., Shimizu, T., Otsubo, M., Nagata, S., *et al.*: 2008, *Solar Phys.* **249**, 167.

Ulmschneider, P., Zähringer, K., Musielak, Z.E.: 1991, *Astron. Astrophys.* **241**, 625.

Vecchio, A., Cauzzi, G., Reardon, K.P., Janssen, K., Rimmele, T.: 2007, *Astron. Astrophys.* **461**, L1.

Vecchio, A., Cauzzi, G., Reardon, K.P.:2009, *Astron. Astrophys.* **494**, 269.